\documentclass[prl,twocolumn,floatfix,footinbib,superscriptaddress,showpacs, showkeys,notitlepage]{revtex4-1}
\usepackage{graphicx}
\usepackage{amssymb}
\usepackage[utf8]{inputenc}
\usepackage[english]{babel}
\usepackage[usenames,dvipsnames]{color}
\usepackage{longtable}
\usepackage{array}
\usepackage{multirow}
\usepackage{ulem}
\usepackage{hyperref}

\usepackage{amsmath}
\newcommand{\Tr}{\text{Tr}}
\newcommand{\vc}[1]{\mathbf{#1}}

\begin{document}
\title{Critical Exponents of Strongly Correlated Fermion Systems from Diagrammatic Multi-Scale Methods}
\author{Andrey~E.~Antipov}\email{aantipov@umich.edu}
\affiliation{Max Planck Institute for Chemical Physics of Solids, N\"othnitzer Stra\ss{}e 40, 01187 Dresden, Germany}
\affiliation{Max Planck Institute for the Physics of Complex Systems, N\"othnitzer Stra\ss{}e 38, 01187 Dresden, Germany}
\affiliation{Department of Physics, University of Michigan, Ann Arbor, Michigan 48109, USA}

\author{Emanuel Gull}\email{egull@umich.edu}
\affiliation{Department of Physics, University of Michigan, Ann Arbor, Michigan 48109, USA}
\author{Stefan Kirchner}\email{kirchner@pks.mpg.de}
\affiliation{Max Planck Institute for Chemical Physics of Solids, N\"othnitzer Stra\ss{}e 40, 01187 Dresden, Germany}
\affiliation{Max Planck Institute for the Physics of Complex Systems, N\"othnitzer Stra\ss{}e 38, 01187 Dresden, Germany}

\date{\today}

\begin{abstract}
Self-consistent dynamical approximations for strongly correlated fermion systems are particularly successful in capturing the dynamical competition of local correlations. In these, the effect of spatially extended degrees of freedom is usually only taken into account in a mean field fashion or as a secondary effect. As a result, critical exponents associated with phase transitions have mean field character. Here, we demonstrate that diagrammatic multi-scale methods anchored around local approximations are indeed capable of capturing the non mean-field nature of the critical point of the lattice model encoded in a non-vanishing anomalous dimension, and to correctly describe the transition to mean field like behavior as the number of spatial dimensions increases.
\end{abstract}


\maketitle

Lattice models of correlated fermions  appear in a wide variety of physical systems, from condensed matter, where they are used to study low-energy models of, {\it e.g.}, transition metals and intermetallic rare earth and actinide compounds, to quantum chemistry and quantum chromodynamics. Many of these systems exhibit a plethora of phases that arise out of the dynamic competition of the underlying degrees of freedom. Understanding the possible orders that characterize these phases and the various ways they can disappear is of primary interest in current many-body physics.
  
The universal theory for the continuous melting of order at finite temperature applies to the `scaling' regime in the immediate vicinity of the transition temperature $T_c$. Relating it to microscopic models of lattice fermions requires knowledge of the underlying universality class.
Even for classical systems, identifying this universality class is difficult and relies, in general, on numerical techniques.
The main difficulty lies in the divergence of scales as $T_c$  is approached.
Large systems paired with a careful finite size analysis are needed and a finite size ansatz and cluster updates or similar techniques are required to overcome the concomitant critical slowing down. For quantum systems, such analyses can be done on bosonic and spin systems \cite{Evertz1993,Capogrosso2008} where lattice Monte Carlo and series expansion techniques allow access to large enough systems to perform a reliable finite size scaling.

The situation for correlated fermionic lattice systems important to condensed matter is different. Analytical solutions only exist for very special setups or weak and strong coupling limits. Semi-analytical infinite partial summation techniques \cite{Bickers2004} can be applied but are in general uncontrolled. Approximate numerical approaches (among them the dynamical mean field theory (DMFT) \cite{Georges1996}, which does not include any non-local correlations), are known to yield inaccurate critical behavior in two and three dimensions \cite{Freericks2003, Anders2011}. Similarly, large-scale numerical efforts, e.g. diagonalization, lattice quantum Monte Carlo, or large diagrammatic simulations have not yet reached the system sizes necessary to perform a reliable finite size scaling analysis at a critical point. For general systems and parameter regimes, the exponential scaling of these techniques (due to an exponential growth of the Hilbert space or the fermionic sign problem) makes it prohibitively expensive to directly simulate large systems.

Diagrammatic multi-scale approaches \cite{Toschi2007,Rubtsov2008,Slezak2009,Taranto2013} offer an elegant potential remedy to this problem: the difficult correlated part of the system is solved using a non-perturbative many-body method, whereas `easier', `weakly correlated' parts of the problem are subsequently tackled in a perturbative scheme. As perturbation theory generically fails near criticality, it is a priori unclear if such a method can accurately describe the behavior near critical points.
A first application to the critical properties of the antiferromagnetic phase transition in the 3d Hubbard model in terms of the dynamical vertex approximation (D$\Gamma$A) \cite{Toschi2007} has appeared in \cite{Rohringer2011}.
This calculation, however, was not fully self-consistent at the two-particle level and required an additional rescaling of
the two-particle vertex functions of  D$\Gamma$A introduced in \cite{Held2008,Katanin2009}, resulting in a Gaussian description of the spatial fluctuations characterized by anomalous dimension $\eta=0$ \cite{Rohringer2011}.
The 3d Hubbard model in the infinite interaction strength limit undergoes a continuous classical ({\it i.e.} finite temperature) phase transition which is in the universality class of the $\phi^4$-theory. The spatial correlations at criticality thus are characterized by $\eta \approx 0.03$ \cite{Holm1993} close to the Gaussian limit $\eta=0$, which makes it difficult to test if the correct non-mean field spatial behavior of the interacting system is captured faithfully.
 
In this letter we present results for the phase diagram and criticality of a simple fermionic model obtained from a multi-scale method that is able to describe non-Gaussian critical behavior. 
Our multi-scale method is the single-site DMFT combined with a dual fermion (DF) approach~\cite{Rubtsov2008}, a diagrammatic expansion in terms of the reducible DMFT vertices. This scheme has previously been shown to have a faster convergence than regular lattice perturbation theories, as it has a small parameter in both the weak- and strong-coupling regimes \cite{Rubtsov2009, Hafermann2009}.  The starting point of the dual fermion expansion is the non-perturbative DMFT expression for the lattice Green's function \cite{Rubtsov2008} and it does not require an \textit{a posteriori} renormalization of vertex functions. We focus on the half-filled Falicov-Kimball (FK) \cite{Falicov1969} model in two, three and four dimensions.
This model arises in various contexts ranging from binary alloys to charge order in intermediate-valence systems and is frequently used to study phase transitions \cite{Freericks2003}. We choose this model because of its continuous phase transition which at infinite interaction strength is described by a $\phi^4$-theory with a $Z_2$ order parameter \cite{Kennedy1986} which leads to strongly non-Gaussian spatial fluctuations in 2d encoded in $\eta=1/4$; and because of its numerical tractability at the DMFT level \cite{Freericks2003}.

Our results show that non-Gaussian critical behavior in the vicinity of a phase transition and critical exponents can be extracted reliably from diagrammatic multi-scale simulations, thereby enabling the use of these methods in more general and more demanding contexts.

\paragraph{Model and Method}
We consider the FK model in $d$ dimensions with two types of electrons: `heavy', `fixed' $f$ electrons characterized by a level energy $\epsilon_d$ and the operators $f^{(\dagger)}$ and `light', `mobile' conduction electrons described by the operators $c^{(\dagger)}$ with the dispersion of the hypercubic lattice $\epsilon_{\bf k} = -2t\sum_j^d\cos k_j$:
\begin{equation} \label{fk_hamilt}
\hat H = \sum_{\vc{k}} \varepsilon_{\vc{k}} c^\dagger_{\vc{k}} c_{\vc{k}} + \sum_i \varepsilon_d f^\dagger_i f_i + U \sum_i c^\dagger_i c_i f^\dagger_i f_i,
\end{equation}
where $U$ is the interaction strength between $c$ and $f$ electrons. Throughout this paper we will use $t=1$ as unit of energy.
 
Within DMFT,  the solution of the lattice problem is approximated by a numerically tractable local impurity problem with an effective self-consistently determined hybridization function $\Delta_{\omega}$. The action of the  lattice problem (\ref{fk_hamilt}) in imaginary time is $S = \sum S^{\mathrm{imp}}[c,c^*,n_f] - \sum_{\omega\vc{k}} \left( \Delta_{\omega} - \varepsilon_{k} \right) c^*_{\omega \mathbf{k}} c_{\omega \mathbf{k}}$, where the impurity action $S^{\mathrm{imp}}$ reads:
\begin{multline}
\label{fk_imp_action}
S^{\mathrm{imp}}[c,c^*,n_f]=\sum_{\omega}\left(\Delta_\omega - i\omega - \mu\right)c^*_{\omega}c_{\omega}+\\+\beta (\varepsilon_d - \mu ) n_f + U n_f \sum_\omega c^{*}_{\omega}c_{\omega}.
\end{multline}
Here $c_{\omega},c^{*}_{\omega}$ are conduction electron fields at the impurity site, $n_f = 0,1$ is a classical variable representing the local occupation number of $f$-electrons and $\beta=T^{-1}$ is the inverse temperature. The solution of the single impurity problem (\ref{fk_imp_action}) can be obtained analytically \cite{Brandt1989}, and in the infinite coordination number limit DMFT yields exact expressions \cite{Metzner1989,Zlatic1990}  for the Green's and vertex functions. We denote the Green's function of the interacting impurity problem as $g_{\omega} = -\langle c_{\omega} c^{*}_{\omega}\rangle$. The hybridization function $\Delta_{\omega}$ is set by taking $g_{\omega}$ equal to the local part of the lattice Green's function $g_{\omega} =  \sum_{\vc{k}} G_{\omega\vc{k}}^{\mathrm{DMFT}}$. In the FK model, the coupling of the itinerant $c$-electrons to the classical $f$-electrons leads to a separation of the static and dynamic response functions \cite{Freericks2000a}, and to a simplified two-frequency (rather than three-frequency) dependence of the two-particle Green's function $\chi_{\omega\omega'} = \langle c_{\omega} c^*_{\omega} c_{\omega'} c^*_{\omega'}\rangle$.  Moreover, the higher order (three-, four-particle \dots) impurity vertex functions vanish in the particle-hole symmetric regime of the model (see derivation in supplementary material). In the finite dimensional case considered here, DMFT is an approximation and yields a set of mean-field exponents for the charge-ordering transition \cite{Freericks2003}.

\begin{figure}[t]
\includegraphics[width=\columnwidth]{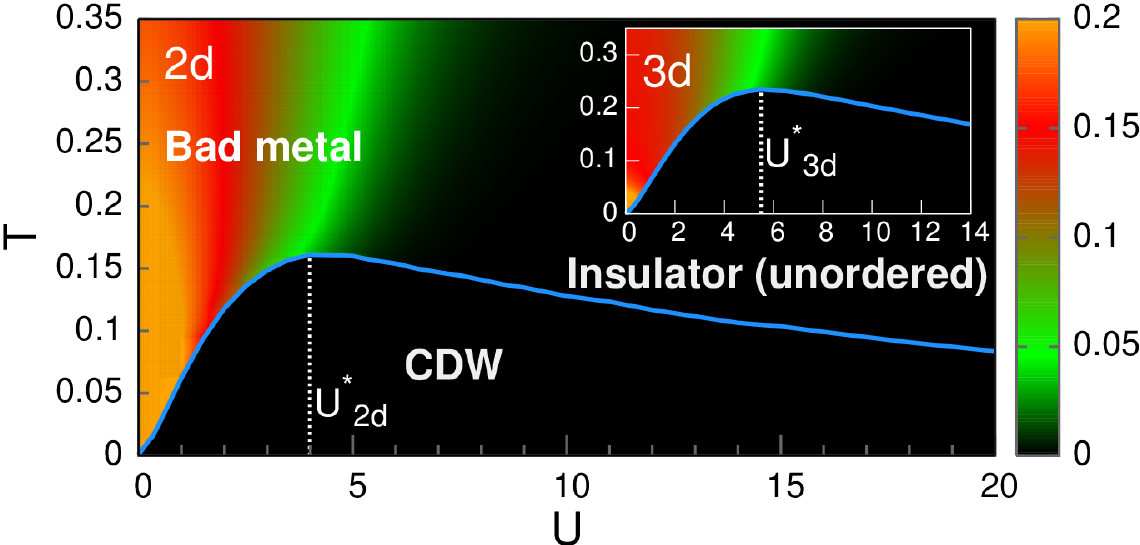}
\caption{Estimate for the Fermi energy density of states $A(0)\sim-\beta G_{\mathrm {loc}}(\beta/2)/\pi$ of $c$-electrons at $\varepsilon = \varepsilon_F$ in $2d$, (inset: $3d$) as a function of $U$ and $T$ at half-filling. Solid line $T_c(U)$ indicates a phase transition to the checkerboard ordered phase (CDW). Above $T_c$ the model shows a crossover from a metallic phase to the isotropic insulator.}
\label{fig1}
\end{figure}

We then apply our multi-scale dual fermion method to obtain non-local correlations. Following Ref.~\onlinecite{Rubtsov2008} we introduce a set of dual fermions $\xi,\xi^*$ and integrate out the original degrees of freedom to rewrite the action of the original lattice problem (\ref{fk_hamilt}) as  
\begin{align}\label{dual_action}
\tilde S = -\!\!\!\sum_{\omega\vc{k}}\xi^{*}_{\omega\vc{k}}\!\!\left[ \tilde G_{\omega\vc{k}}^{0}\right]^{-1}\!\!\!\!\xi_{\omega\vc{k}} - \frac{1}{2} \sum_{\omega\omega' j} \gamma_{\omega\omega'} \xi^*_{\omega i}\xi_{\omega i}\xi^*_{\omega' i}\xi_{\omega' i}.
\end{align}
Here $\gamma_{\omega\omega'} =  g_{\omega}^{-2} \left[ \chi_{\omega\omega'} - g_{\omega}g_{\omega'} \right]  g_{\omega'}^{-2}$ is the reducible two-particle vertex of the DMFT impurity problem (\ref{fk_imp_action}) and $\tilde G_{\omega\vc{k}}^{0} = G^{\mathrm{DMFT}}_{\omega\vc{k}} - g_{\omega}$ denotes the bare dual fermion Green's function.

The dual fermion self-energy is constructed from the reducible DMFT impurity vertex functions.
The summation of all of vertex diagrams generates the exact lattice solution but cannot be performed in practice. At this point we therefore resort to  a ladder approximation and express the connected part of the dual two-particle Green's function as
\begin{equation}\label{bs_static}
\tilde\Gamma_{\omega\omega'}(\vc{q}) = \gamma_{\omega\omega'} -  \sum_{\omega''}\gamma_{\omega\omega''}  \sum_{\vc{k}} \tilde G_{\omega''\vc{k}}\tilde G_{\omega''\vc{k+q}} \tilde\Gamma_{\omega''\omega'}(\vc{q}).
\end{equation}
Its zero-order approximation ($\tilde G = \tilde G^0$) results in DMFT expressions for both the Green's function and the charge susceptibility of the system (see Ref.~\onlinecite{Hafermann2012} and supplementary information). The dual fermion self-energy $\tilde \Sigma_{\omega\vc{k}}$ is obtained from $\tilde\Gamma_{\omega\omega'}(\vc{q})$ via  \begin{equation}\label{dual_selfenergy}
\tilde\Sigma_{\omega\vc{k}} = \sum_{\vc{q}} \tilde\Gamma_{\omega\omega}(\vc{q}) \tilde G_{\omega\vc{k+q}}.
\end{equation}

\begin{figure}[t]
\begin{center}
\includegraphics[width=\columnwidth]{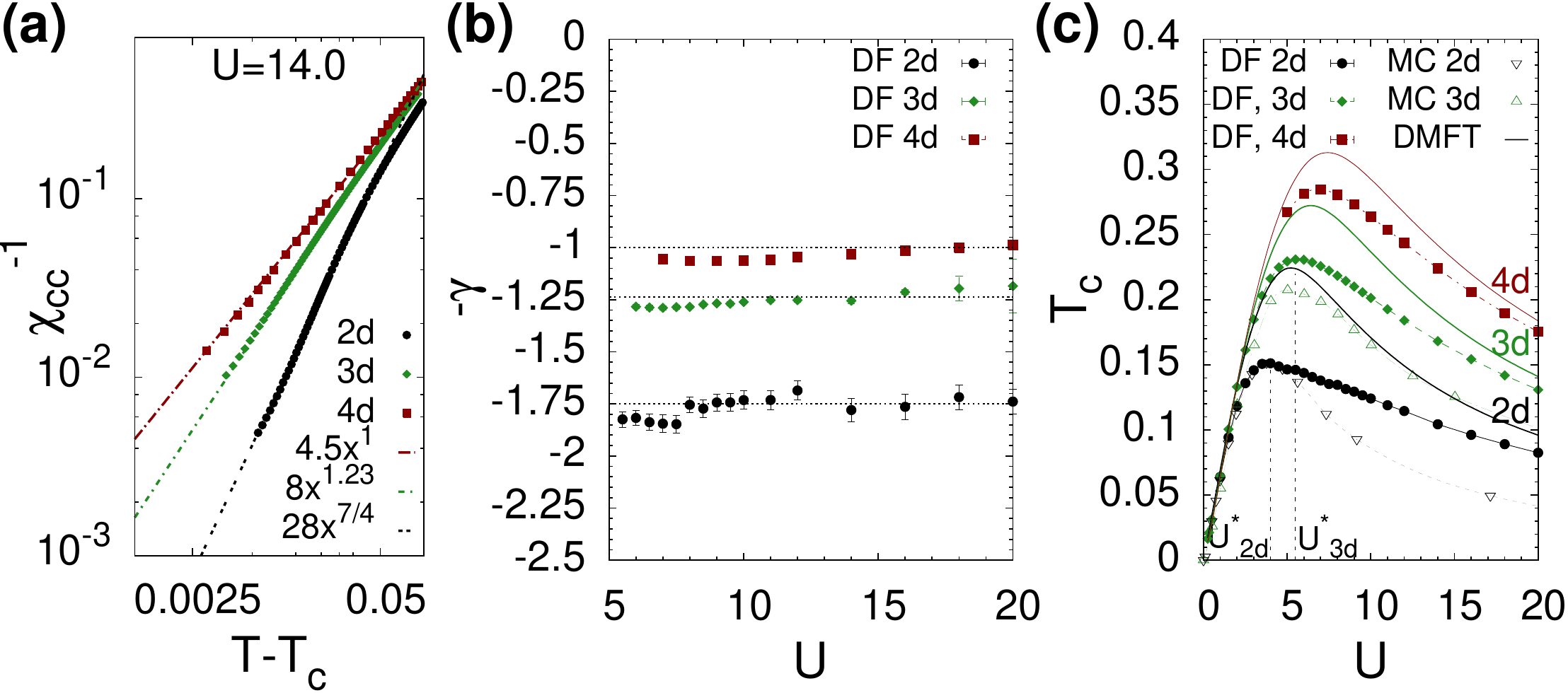}\end{center}\vspace*{-1.5em}
\caption{(a) Inverse of static c-electron charge susceptibility in $d=2,3,4$ as a function of $T-T_c$ at $U=14$. (b) Critical exponent $\gamma$ in $d=2,3,4$ at different values of $U$  for $U > U^*$. Lines: prediction for Ising universality class, $\gamma=1.75$ in $2d$, $1.24$ in $3d$, $1$ in $4d$. Error bars represent regression errors. (c) Critical temperatures of the charge ordering transition in $d=2,3,4$ for different values of $U$. Solid lines: DMFT, dimensions indicated on graph. Lines with heavy symbols: ladder DF. Light symbols: Monte-Carlo of Refs \cite{Maska2006,Zonda2009}. }
\label{fig2}
\end{figure}

Eq.~(\ref{bs_static}) and (\ref{dual_selfenergy}) are solved self-consistently. This allows for non-mean field critical exponents, whereas a finite subset of the diagrams does change $T_c$ but only results in mean field exponents (see Supplementary and Ref. \cite{Brener2008}).
The full lattice Green's function $G_{\omega\vc{k}}$ and the lattice two-particle vertex $\Gamma_{\omega\omega'\vc{k}\vc{k'}}(\vc{q})$ can be obtained from $\tilde G_{\omega\vc{k}}$ and $\tilde \Gamma_{\omega\omega}(\vc{q})$ via $G_{\omega\vc{k}} = (\Delta_{\omega} - \varepsilon_{\vc{k}})^{-1} +  g_{\omega}^{-2} (\Delta_{\omega} - \varepsilon_{\vc{k}})^{-2} \tilde G_{\omega\vc{k}}$ and 
 $\Gamma_{\omega\omega'\vc{k}\vc{k'}}(\vc{q}) = L_{\omega\vc{k}}L_{\omega\vc{k+q}}L_{\omega'\vc{k'}}L_{\omega'\vc{k'+q}} \tilde \Gamma_{\omega\omega'}(\vc{q})$, where 
 $L_{\omega\vc{k}} = [ 1 - \tilde \Sigma_{\omega\vc{k}} g_{\omega} ]^{-1}$. 
The static lattice $c$-electron charge susceptibility becomes $\chi_{cc}(\vc{q}) = - T\sum_{\omega\vc{k}} G_{\omega\vc{k}}G_{\omega\vc{k+q}} +  T\sum_{\omega\omega'}\sum_{\vc{k}\vc{k'}} G_{\omega\vc{k}}G_{\omega\vc{k+q}} \Gamma_{\omega\omega'\vc{k}\vc{k'}}(\vc{q})G_{\omega'\vc{k'}}G_{\omega'\vc{k'+q}}$.

\paragraph{Results}
We present the phase diagram of the half-filled FK model on $2d$ and $3d$ cubic lattices obtained in the isotropic phase of the model in Fig.~\ref{fig1}.
Shown is our estimate for the $c$-electron density of states at $\omega=0$ obtained from the DF calculation as $A(\omega=0)\sim-\beta G_{\mathrm {loc}}(\beta/2)/\pi$. We find three different phases: below $T_c$ (indicated by the solid line) a checkerboard charge-ordered state exists for all values of $U$; above $T_c$ the isotropic metallic non-Fermi liquid state \cite{Si1992} for low  $U$ and a disordered Mott-like insulating state for large $U$ are separated by a crossover with a characteristic energy scale $U^*$ indicated by a dashed line in Fig. \ref{fig1}. At $U>U^*$ a gap in the spectrum is present for all temperatures and the kinetic energy of the conduction electrons is small compared to the potential energy and the phase transition is of second order. At $U<U^*$ Monte Carlo simulations on finite-size clusters \cite{Maska2006, Zonda2009} report first-order coexistence, whereas DMFT finds an unconventional continuous phase transition \cite{Freericks2000c, VanDongen1992}.

\begin{figure}[t]
\begin{center}
\includegraphics[width=\columnwidth]{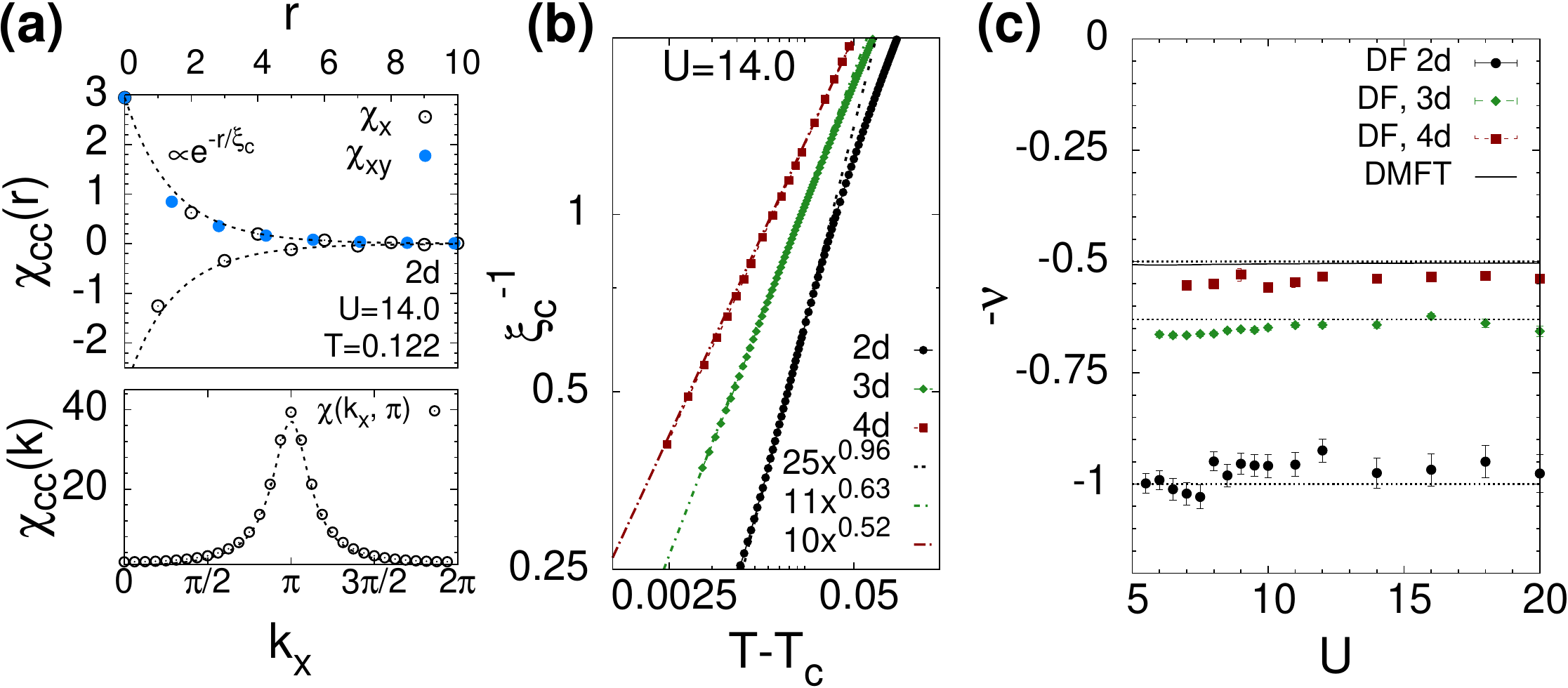}\end{center}\vspace*{-1.5em}
\caption{
(a) $\chi_\text{cc}$ as a function of distance (top) along the $x$ and $xy$ direction, for $U=14$ and $T=0.122$. Dashed line: exponential envelope of $\chi_\text{cc}$. Bottom: data and fit to Lorentzian in momentum space for same parameters. (b) Inverse correlation length $\xi^{-1}$ and power law fit as a function of $T-T_c$ at  $U=14$, for $d=2,3,4$. (c) Correlation length exponent $\nu$ for $d=2,3,4$ along with theoretical predictions for the Ising model (dashed lines) and DMFT results (solid line) as a function of $U$ for $U > U^*$.}
\label{fig3}
\end{figure}

In Figs.~\ref{fig2} and \ref{fig3} we analyze the critical behavior of the continuous transition at $U>U^*$. In Panel \ref{fig2}a we show the inverse of the static checkerboard $\vc{q} = (\pi,\pi,\hdots)$ $c$-electron susceptibility $\chi_\text{cc}$ on approach to the charge ordering phase transition in $2$, $3$ and $4$ dimensions as obtained from our DF  calculations. The divergence of $\chi_\text{cc}$ indicates a phase transition to the checkerboard charge-ordered phase, which occurs at half-filling. We find power law behavior for about two decades in $T-T_c$ and $\chi_{cc}^{-1}$, allowing us to fit the critical exponent $\gamma$ with good confidence. Panel \ref{fig2}b shows the evolution of $\gamma$ as a function of $U$ along the second order line. We observe a significant dimensionality dependence of $\gamma$ and an agreement with the Ising universality class (of the respective dimension) within error bars for the entire range $U > U^*$. Our error bars reflect the least square fitting uncertainty for the exponents. 
In panel~\ref{fig2}c we show the critical temperature of the phase transition in different dimensions as obtained by Monte-Carlo simulations \cite{Maska2006,Zonda2009} (obtained on finite-size clusters), DMFT and dual fermions. Nonlocal correlations in the system lead to a decrease of the critical temperature from the overestimated DMFT values towards the Monte Carlo values. The DF improves on the DMFT for all interaction values studied. The difference between the dual fermion $T_c$ and the true transition temperature is determined by all the diagrams not included in the DF approximation.

In Fig.~\ref{fig3}a we show the spatial dependence of $\chi_\text{cc}$ along the $x$ (open black symbols) and the $xy$ (solid blue symbols) directions. The spatial modulation of $\chi_\text{cc}$ shows checkerboard ordering, indicated by a change of sign of susceptibility in $x$-direction. The data is consistent with an exponential decay corresponding to a Lorentzian in momentum space (lower panel), so that we can extract a correlation length $\xi$. Fig.~\ref{fig3}b shows the critical behavior of $\xi$ as a function of $T$ at $U=14$, which in agreement with Fig.~\ref{fig2}b shows a strong dimensionality dependence. The corresponding critical exponent $\nu$, plotted in Fig.~\ref{fig3}c as a function of $U$, shows reasonable agreement with the Ising exponents for the entire range of $U$. 

As all critical properties are obtained from the effective action and its derivatives, scaling relations are expected to hold. From $\gamma$ and $\nu$ one therefore can extract the  value of $\eta = 2 - \gamma/\nu$ ($0.25$ in 2d, $0.03$ in 3d), highlighting the correct description of non-Gaussian spatial fluctuations at the phase transition. This result is more general: in the Hubbard model, vertex functions will acquire additional dynamic components, which are absent in the Falicov-Kimball model. However, due to the decoupling of statics and dynamics at a classical phase transition \cite{Popov1983}, critical exponents only depend on the lattice dimension and the order parameter symmetry and are insensitive to the dynamics of the underlying microscopic model \cite{Wilson1972}. Correct critical exponents for the Hubbard model may be obtained from an analogous calculation using only static vertex components. In contrast, non-universal quantities (e.g. $T_c$) will depend on dynamic vertex parts and require a more extensive analysis. 

\begin{figure}[bt]
\begin{center}
\includegraphics[width=\columnwidth]{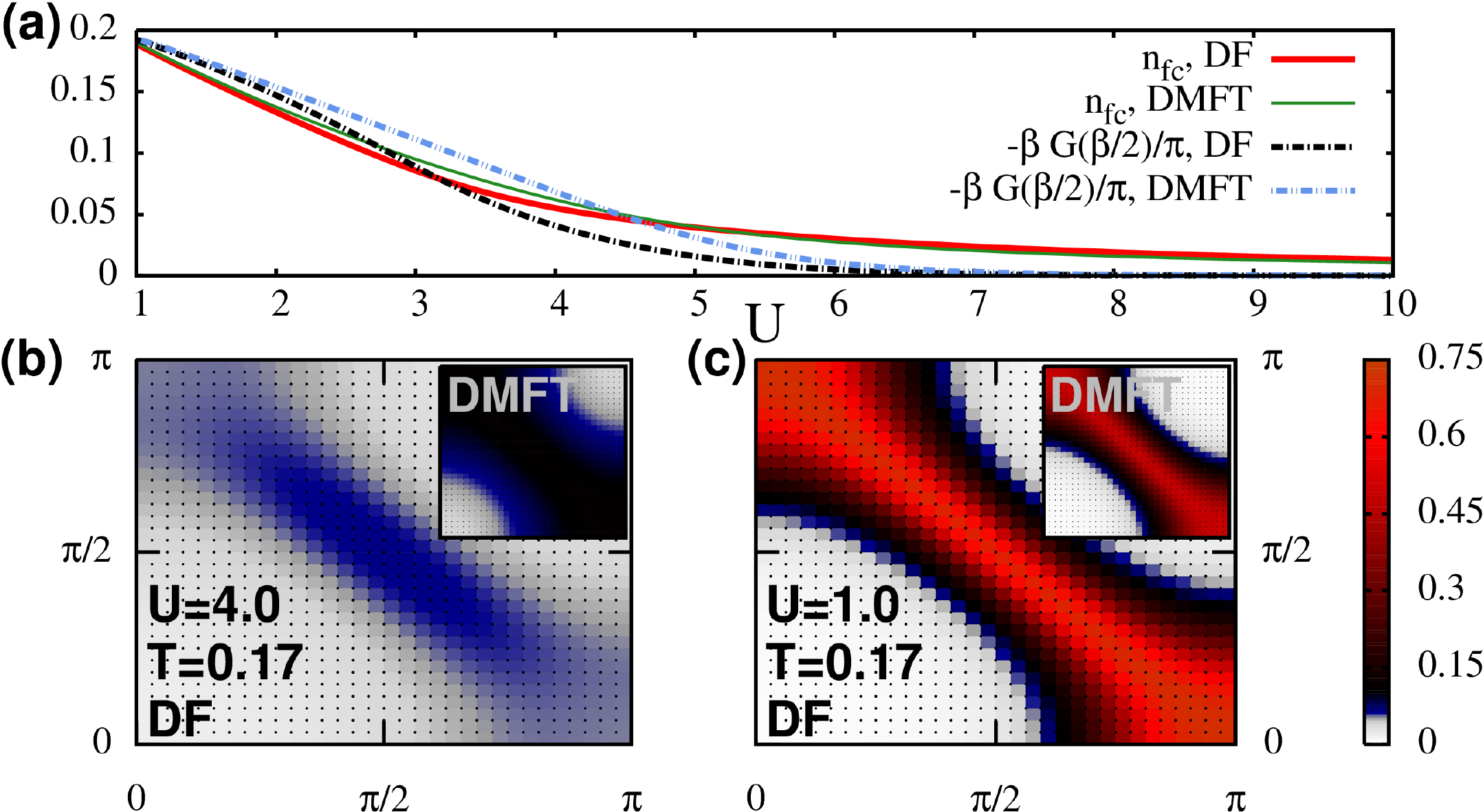}\end{center}
\caption{(a) $fc$ occupancy $n_{fc}$ (solid lines) and local density of states (DOS) estimate $-\beta G_\text{loc}(\beta/2)/\pi$ (dashed lines) in $2$d as a function of $U$ in DMFT and DF, both showing a smooth crossover from metal to Mott-like insulator. Bottom: Momentum-resolved $-\beta G_k(\beta/2)$ for $U=4.0$ (b) and $U=1.0$ (c), as obtained in DF (main panel) and DMFT (insets). Black dots indicate the mesh points on which DF and DMFT are evaluated.
}
\label{fig4}
\end{figure}

We now turn to a discussion of the electronic properties  that underlie this behavior. 
Fig.~\ref{fig4} panel (a) shows the local $fc$ occupancy $n_{fc} = \langle c^\dagger c f^\dagger f \rangle_{\mathrm{loc}}$ in $2$ dimensions and our estimate for the local density of states as a function of $U$ near the crossover between metallic and insulating states at $T>T_c$.
DMFT calculations \cite{VanDongen1992} showed continuous crossover behavior and no hysteresis for both quantities. Non-local correlations do not change this behavior qualitatively, but we find that the crossover scale $U^*$ is slightly smaller than in DMFT. We also find, consistent with expectation, that deviations from the DMFT results are largest near the point where the curve is closest to the phase transition to the ordered state and critical phenomena become important. Away from this region, DF and DMFT results are remarkably similar and non-local correlations appear less important.

The momentum-resolved spectral function estimates shown in panel (b) and (c) of Fig.~\ref{fig4} highlight this further. DF (main panels) and DMFT (insets) DOS at the Fermi energy are shown as a function of momentum for $U=4$ and $U=1$. The temperature is chosen to be just above $T_c$ for $U=4$. The weak correlation regime deep within the phase at $U=1$ is well captured by the momentum-independent DMFT self energy. Near $T_c$ at $U=4$, the deviations between the DF and DMFT solutions become substantial. DMFT shows a comparatively large and $k$-independent spectral function in a large area around the non-interacting Fermi surface. DF, in contrast, exhibits a much narrower and non-trivial momentum dependent spectral function that suppresses the density of states near the $(\pi,0)$ and $(0,\pi)$ points but not near the zone diagonal $(\pi/2,\pi/2)$ points. This behavior is expected because a divergence in the two-particle propagator feeds back into the single-particle Green's function through Eq.~(\ref{dual_selfenergy}). The effect is reminiscent of the momentum-selective pseudogap regime in the Hubbard model, where both short ranged correlations contained in, {\it e.g.}, cluster DMFT \cite{Gull2009} or long ranged correlations contained in DF and D$\Gamma$A calculations \cite{Rubtsov2009, Hafermann2009, Katanin2009} result in qualitatively similar behavior.

\begin{figure}[tb]
\begin{center}
\includegraphics[width=\columnwidth]{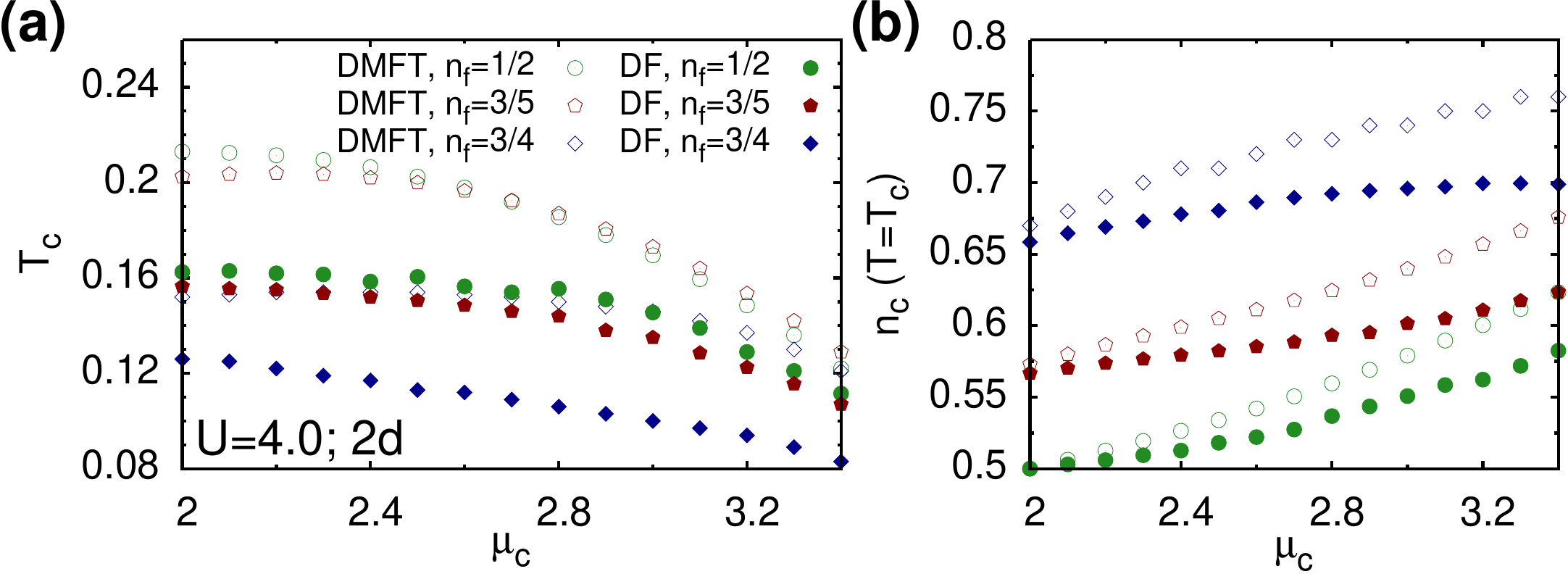}\end{center}
\caption{(a) Transition temperature $T_c$ and (b) average occupation number of $c$-electrons at $T=T_c$ as a function of $c$-electron chemical potential in 2d at $U=4$ at an average $f$ electron occupation number of $n_f = 1/2; 3/5; 3/4$. DMFT results: open symbols. DF: filled symbols.
}
\label{fig5}
\end{figure}

The charge order phase transition in this model persists away from half filling \cite{Lemanski2002}. Fig.~\ref{fig5}a shows $T_c$ as a function of c-electron chemical potential for three different $f$-electron concentrations. We find that doping suppresses $T_c$ both in DMFT (open symbols) and DF (filled symbols), with the DF $T_c$ always below DMFT. Panel (b) shows the evolution of the c-electron density as a function of c-electron chemical potential at $T_c$. The large difference between DF and DMFT results is a consequence of correlation effects absent in DMFT. 

\paragraph{Summary} In conclusion, we have applied the `dual fermion' diagrammatic multi-scale method to an interacting quantum many-body system. We demonstrated that our dual fermion ladder approximation recovers the exact interacting critical behavior in $d=2,3,4$ dimensions and that diagrammatic multi-scale methods can yield accurate results for the criticality of interacting fermionic lattice models. We also showed that the method provides valid results away from particle-hole symmetry. It may therefore be able to provide accurate predictions for more complicated (and in particular frustrated) models \cite{Li2014}, where no reliable theoretical tools are available so far, as well as for transport properties near and far from equilibrium \cite{Munoz2013, Zlatic2012}. It also motivates the use of multi-scale methods to gain insight into critical phenomena that are far less clear, such as the superconducting transition in the cuprates \cite{Mishra2014} and the quantum melting of magnetism in certain heavy fermion compounds \cite{Si2010}, where many continuous quantum phase transitions appear to be beyond the traditional classification scheme \cite{Si2014,Stockert2012}.
 
\paragraph{Acknowledgments} We thank Romuald Lemanski, Gang Li, Pedro Ribeiro and Alexey N. Rubtsov for fruitful discussions. We used the TRIQS package for post-processing data \footnote{TRIQS, \url{http://ipht.cea.fr/triqs/}}. EG acknowledges support of the Simons foundation.
\bibliographystyle{apsrev4-1}
\bibliography{fkbib}

\cleardoublepage

\newpage
\clearpage
\appendix

\begin{widetext}
\begin{center} {\large \bf Critical Exponents of Strongly Correlated Fermion Systems from Diagrammatic Multi-Scale Methods: Supplementary Materials }  

\bigskip 
Andrey E. Antipov,$^{1,2,*}$ Emanuel Gull,$^{3,\dagger}$ and Stefan Kirchner$^{1,2,\ddagger}$\\
\medskip 
{\it \small
$^1$Max Planck Institute for Chemical Physics of Solids, N\"othnitzer Stra\ss{}e 40, 01187 Dresden, Germany\\
$^2$Max Planck Institute for the Physics of Complex Systems, N\"othnitzer Stra\ss{}e 38, 01187 Dresden, Germany\\
$^3$Department of Physics, University of Michigan, Ann Arbor, Michigan 48109, USA
}
\end{center}

\setcounter{figure}{0}   \renewcommand{\thefigure}{S\arabic{figure}}
\setcounter{equation}{0} \renewcommand{\theequation}{S.\arabic{equation}}
\setcounter{section}{0} \renewcommand{\thesection}{S.\Roman{section}}
\renewcommand{\thesubsection}{S.\Roman{section}.\Alph{subsection}}
\makeatletter
\renewcommand*{\p@subsection}{}  
\makeatother
\renewcommand{\thesubsubsection}{S.\Roman{section}.\Alph{subsection}-\arabic{subsubsection}}
\makeatletter
\renewcommand*{\p@subsubsection}{}  
\makeatother
\renewcommand*{\citenumfont}[1]{S#1}
\renewcommand*{\bibnumfmt}[1]{[S#1]}
\newcommand{\citesup}[1]{\citetext{S\citealp{#1}}}

In this supplement we
\begin{itemize}
\item Refine the analytical solution of the single impurity DMFT problem, give expressions for the two-particle reducible vertex function and provide evidence for the vanishing of the reducible three-particle vertex function of the impurity problem.
\item Solve the dual perturbation theory in the ladder approximation (S2).
\item Show the equivalence between the ladder approximation for the dual fermion vertex in the bare approximation for the Green's function and the standard DMFT vertex expression. (S3)
\item Compare static lattice susceptibility obtained by a ladder and the second-order dual fermion approximations. (S4)
\end{itemize}

\section{S1. Green's functions and reducible vertices of the effective DMFT impurity problem}
In this section we briefly repeat the solution to the single-impurity problem (Eq. \ref{fk_imp_action} in the main text) previously given in Refs.~\citesup{supBrandt1989} and \citesup{supFreericks2003} and provide the expressions for the single-, two- and three-particle Green's functions and corresponding reducible vertex functions. The action of the problem reads:
\begin{equation}
\label{S:fk_imp_action}
S^{\mathrm{imp}}[c,c^*,n_f]=\sum_{\omega}\left(\Delta_\omega - i\omega - \mu\right)c^*_{\omega}c_{\omega}+\beta (\varepsilon_d - \mu ) n_f + U n_f \sum_\omega c^{*}_{\omega}c_{\omega},
\end{equation}
where $c,c^*$ labels conduction electrons and $n_f=0,1$ the occupation number of heavy electrons. $\Delta_{\omega}$ is the hybridization function. The partition function is 
$Z = \Tr_{n_f} \int \mathcal{D}c\mathcal{D}c^* \exp(-S^{\mathrm{imp}}[c,c^*,n_f]) = Z^{n_f=0} + Z^{n_f=1}$, where $Z^{n_f} = \prod_\omega \int  \mathcal{D}c_{\omega}\mathcal{D}c^*_{\omega} \exp \left[ (K_{\omega}^{n_f})^{-1}  c^{*}_{\omega} c_{\omega}  \right]$ and 
\begin{equation}
K^{n_f=0,1}_{\omega} = \left[i\omega + \mu - \Delta_\omega - U n_f \right]^{-1}.
\end{equation}
An average value of an operator $O$ with the action (\ref{S:fk_imp_action}) is represented as a weighted average of two Gaussian problems:
\begin{equation}\label{S:generic_average}
\langle O \rangle_{\mathrm{imp}} = w_0 \langle O \rangle_0 + w_1 \langle O \rangle_1,
\end{equation}
where $\langle O \rangle_{f = 0,1} = Z_j^{-1} \int \mathcal{D}c\mathcal{D}c^* O \exp(-S^{\mathrm{imp}}[c,c^*,w=f]) $ and $w_{n_f} = Z^{n_f} / Z$.

The complete interacting Green's function of the impurity reads:
\begin{equation}\label{S:imp_gf}
g_{\omega} = -\langle c_\omega c^*_\omega\rangle_{\mathrm{imp}} = w_0 K^{0}_{\omega} + w_1 K^{1}_{\omega}
\end{equation}
The self-energy of the impurity is 
\begin{equation}\label{S:imp_sigma}
\Sigma_{\omega} = (K^0_\omega)^{-1} - g_\omega^{-1} = w_1 U \frac{K^1_\omega}{g_{\omega}}.
\end{equation}
where we used $K^{1}_{\omega} -  K^{0}_{\omega} = U K^{0}_{\omega} K^{1}_{\omega}$.

The two-particle Green's function of the impurity is
\begin{equation}\label{S:2pgf}
\chi_{1234} = \langle c_{1} c^{\dagger}_{2} c_{3} c^\dagger_{4} \rangle_{\mathrm{imp}} \stackrel{\ref{S:generic_average}}{=} \sum_{n_f = 0,1} w_{n_f} \langle c_{1} c^{\dagger}_{2} c_{3} c^\dagger_{4} \rangle_{n_f} = \sum_{n_f = 0,1} w_{n_f} (\delta_{12}\delta_{34} K^{n_f}_{1}K^{n_f}_{3} - \delta_{23}\delta_{14} K^{n_f}_{1}K^{n_f}_{3}),
\end{equation}
where the shorthand notation $1 = i\omega_1, 2 = i\omega_2, \hdots$ was used. Since $\langle \rangle_{n_f}$ is an average over a Gaussian ensemble Wick's theorem applies. The frequency dependence of the two-particle Green's function is simplified to only two independent frequencies.

The reducible vertex function is the connected part of the Green's function. For each of the components obtained by Wick's theorem it reads 
\begin{subequations}
\begin{align}
\bar\gamma_{12} &= g_1^{-1}g_2^{-1} \bar\Gamma^{(4)}_{12} g_2^{-1}g_1^{-1} \\
\bar\Gamma^{(4)}_{12} &= \sum_{n_f} w_{n_f} K^{n_f}_{1}K^{n_f}_{2} -  g_{1}g_{2}
\end{align}
\end{subequations}
Substituting (\ref{S:imp_gf}) into the last equation we arrive at $\bar \gamma_{12} = w_0 w_1 U^2 (\Lambda_{1}\Lambda_{2})$, where 
\begin{equation}
\Lambda_{\omega} = \frac{K^0_\omega K^1_\omega}{g_{\omega}^2}.
\end{equation}
For the static component $\omega_1 = \omega_2$ both parts of $\chi$ obtained from Wick's theorem will contribute. The complete two-frequency dependent vertex function then reads
\begin{equation}\label{S:gamma4}
\gamma_{\omega\omega'} = w_0 w_1 U^2 (\Lambda_{\omega}\Lambda_{\omega'} - \delta_{\omega\omega'}\Lambda_\omega^2).
\end{equation}
The simplified frequency structure of $\gamma_{\omega\omega'}$ that obeys Wick's theorem leads to a change of the prefactor to $1/2$ (as opposed to $1/4$ in the expression for the dual action [Eq.~\ref{dual_action} in the main text] of the Hubbard model \citesup{supRubtsov2008}). This is understood in the following way: consider the term $\sum_{1234} \gamma_{1234} f^*_1 f_2 f^*_3 f_4$ in the dual action. Using Wick's theorem it reads
\begin{multline}
\sum_{1234} \gamma_{1234} f^*_1 f_2 f^*_3 f_4 = \sum_{13} \gamma_{13} \delta_{12}\delta_{34} f^*_1 f_1 f^*_3 f_3  - \sum_{12} \gamma_{12} \delta_{13}\delta_{24} f^*_1 f_3 f^*_3 f_1 = 2\sum_{13} \gamma_{13} \delta_{12}\delta_{34} f^*_1 f_1 f^*_3 f_3,
\end{multline}
where we relabeled $2\leftrightarrow 3$ in the second sum.

The same considerations as used above for two-particle correlators can be applied for obtaining higher-order Green's function and vertex functions. The three particle Green's function is a sum of $3! = 6$ equivalent contributions that are obtained from each other by reshuffling the indices. We consider one of them, $g_{123} = \sum_{n_f} w_{n_f} K^{n_f}_1 K^{n_f}_2 K^{n_f}_3$. The expression for the third-order vertex function reads:
\begin{subequations}
\begin{align}
\bar\Gamma^{(6)}_{123} &= g_{123} + 2g_{1}g_{2}g_{3} - g_{1} \chi_{23} - g_{2}\chi_{13} - g_{3}\chi_{12} \\
\bar\gamma^{(6)}_{123} &= g_1^{-1} g_2^{-1} g_3^{-1} \bar\Gamma^{(6)}_{123}  g_3^{-1} g_2^{-1} g_1^{-1}.
\end{align}
\end{subequations}
With the previously obtained expressions (\ref{S:imp_gf}) for $g$ and (\ref{S:2pgf}) for $\chi$ and using the following algebraic relations
\begin{gather*}
2w_0^2w_1 -w_0w_1 = - (2w_1^2w_0 -w_0w_1 ) = w_0 w_1 (w_0 - w_1)\\
w_0 - 3w_0^2 + 2w_0^3 = - (w_1 - 3w_1^2 + 2w_1^3) = - w_0 w_1 (w_0 - w_1),
\end{gather*}
one arrives at the following expression:
\begin{equation}
\Gamma^{(6)}_{123} = w_0 w_1 (w_0 - w_1) U^3 K_1^0 K_1^1 K_2^0 K_2^1 K_3^0 K_3^1,
\end{equation}
and thus
\begin{equation}
\gamma^{(6)}_{123} = w_0 w_1 (w_0 - w_1) U^3 \Lambda_1 \Lambda_2 \Lambda_3.
\end{equation}
At half-filling $w_0 = w_1 = 1/2$ the three-particle reducible vertex function therefore vanishes. 

\subsection{S2. Dual fermion perturbation theory in the ladder approximation for the vertex}
In this section we solve the dual perturbation theory in a ladder approximation for the vertex (Eq.~\ref{bs_static} in the main text):
\begin{equation}\label{S:bs_static}
\tilde\Gamma_{12}(\vc{q}) = \gamma_{\omega\omega'} - \sum_{\omega''}\gamma_{\omega\omega''}  \sum_{\vc{k}} \tilde G_{\omega''\vc{k}}\tilde G_{\omega''\vc{k+q}} \tilde\Gamma_{\omega''\omega'}(\vc{q}),
\end{equation}
where $\tilde G_{\omega\vc{k}}$ is the dual fermion Green's function and reducible two-particle vertex $\gamma_{\omega\omega'}$ is introduced in the previous section. Diagrammatically this Bethe-Salpeter equation is represented as:
\begin{equation}\label{S:bs_static_diagram}
\includegraphics[height=64px]{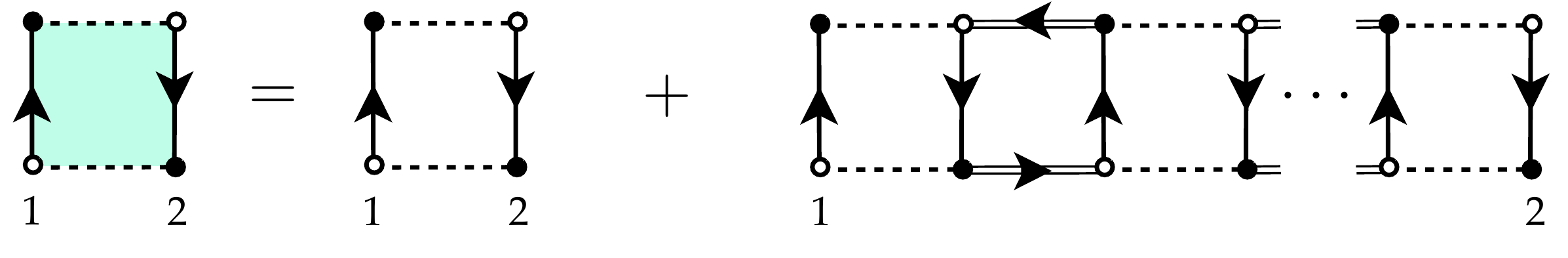},
\end{equation}
where every closed loop of arrows carries a $-1$ sign. 
With the help of the bare dual fermion bubble $\tilde\chi^0_{\omega}(\vc{q}) = -\sum_{\vc{k}} \tilde G_{\omega\vc{k}}\tilde G_{\omega\vc{k+q}}$the equation (\ref{S:bs_static}) is rewritten as:
\begin{equation}\label{S:bs_static_2}
\tilde\Gamma_{12}(\vc{q}) = \gamma_{12} + \sum_3\gamma_{13} \tilde \chi^0_{3}(\vc{q})\tilde\Gamma_{32}(\vc{q}),
\end{equation}
This equation can be solved directly using the matrix inversion, $\tilde \Gamma = [1 - \gamma\tilde\chi^0]^{-1}\gamma$ . It can be further simplified with the help of expression (\ref{S:gamma4}) for the reducible two-particle vertex, $\gamma_{12} = w_0 w_1 U^2 (\Lambda_1\Lambda_2 - \Lambda_1\Lambda_1\delta_{12})$ is used. The procedure is equivalent to the iterative solution of the DMFT irreducible vertex \citesup{supFreericks2003}. Substituting (\ref{S:gamma4}) into (\ref{S:bs_static_2}) one obtains
\begin{equation}\label{S:bs_static_simple2}
\Gamma_{12}(\vc{q}) = \frac{w_0 w_1 U^2 \Lambda_{1} (\Lambda_{2} - \Lambda_{1}\delta_{12})} {1 + w_0 w_1 U^2 \Lambda_{1}^2 \tilde \chi^0_{1}(\vc{q})} + \frac{w_0 w_1 U^2 \Lambda_{1}} {1 + w_0 w_1 U^2 \Lambda_{1}^2 \tilde \chi^0_{1}(\vc{q})} \sum_{3} \Lambda_{3}\tilde \chi^0_{3}(\vc{q})\Gamma_{32}(\vc{q}) 
\end{equation}
Multiplying (\ref{S:bs_static_simple2}) by $\sum_1 \Lambda_1 \tilde \chi^0_{1}(\vc{q})$ and denoting $\phi_{\omega}(\vc{q}) = w_0 w_1 U^2 \Lambda_\omega /\left(1 +  w_0 w_1 U^2 \Lambda_{\omega}^2 \tilde \chi^0_{\omega}(\vc{q})\right)$ and $B(\vc{q},T) = \sum_{\omega} \phi_{\omega}(\vc{q}) \Lambda_{\omega} \tilde \chi^0_\omega(\vc{q})$
one arrives at the algebraic equation for $\xi_{\omega}(\vc{q}) = \sum_{\omega'} \Lambda_{\omega'}\tilde \chi^0_{\omega'}(\vc{q})\Gamma_{\omega'\omega}(\vc{q})$:
\begin{equation}
\xi_\omega (\vc{q}) = B(\vc{q},T) \Lambda_\omega - \phi_{\omega}\Lambda^2_{\omega}(\vc{q}) \tilde \chi^0_\omega(\vc{q}) + B(\vc{q},T) \xi_\omega (\vc{q}),
\end{equation}
which gives
\begin{equation}\label{S:full_static_vertex}
\Gamma_{12}(\vc{q}) =\phi_{1}(\vc{q})\frac{\Lambda_{2} - \Lambda_{1}\delta_{12} + B(\vc{q},T)\Lambda_1\delta_{12} - \phi_2(\vc{q})\Lambda_2^2 \tilde \chi^0_2(\vc{q})}{1-B(\vc{q},T)}.
\end{equation}
\subsection{S3. DMFT vertex relation} 
Eq.~(\ref{S:full_static_vertex}) represents the reducible dual fermion vertex within the ladder approximation (\ref{S:bs_static_diagram}). At the level of DMFT, this expression formally also gives the lattice vertex of the static charge susceptibility (with $\tilde G_{\omega\vc{k}}$ replaced by $G_{\omega\vc{k}}^{\mathrm{DMFT}} - g_{\omega}$) \citesup{supHafermann2012}. To demonstrate the equivalence of formula (\ref{S:full_static_vertex}) to the analogous expression obtained by a skeleton expansion of the impurity self-energy (see Table I of Ref.~\citesup{supFreericks2003}) we first show that the quantities have the same denominators and, consequently, equal critical temperatures and exponents. The divergence of the vertex (\ref{S:bs_static}) occurs when the quantity $B(\vc{q},T) = \sum_\omega \phi_{\omega}$ introduced in the previous section, equals $1$. In Ref. \citesup{supFreericks2003}, the analogous quantity reads:
\begin{subequations}
\begin{align}\label{S:B_DMFT}
B^{\mathrm{DMFT}}(\vc{q},T)=&\sum_{\omega} \frac{w_0 w_1 U^2 g_{\omega}^3 \eta_{\omega}(\vc{q})}
{(1+g_{\omega}\Sigma_{\omega})(1+g_{\omega}(\Sigma_{\omega} - U))\left[ \eta_{\omega}(\vc{q})g_{\omega}(1+g_{\omega}(2\Sigma_{\omega}-U))+ (1+g_{\omega}\Sigma_{\omega})(1+g_{\omega}(\Sigma_{\omega} - U))\right] }\\
\eta_\omega(\vc{q}) =& g_{\omega}\left[-g_{\omega}^{-2} - (\chi^{(0),\mathrm{DMFT}}_{\omega}(\vc{q}))^{-1}\right] =  g_{\omega}\left[-g_{\omega}^{-2} + \left(\sum_{\vc{k}} G_{\omega\vc{k}}^{\mathrm{DMFT}}G_{\omega\vc{k+q}}^{\mathrm{DMFT}}\right)^{-1}\right]
\end{align}
\end{subequations}

We first note that at the DMFT level when $\tilde G_{\omega\vc{k}} = G^{DMFT}_{\omega\vc{k}} - g_{\omega}$, the bare dual bubble can be written as $\tilde \chi^{(0)}_{\omega}(\vc{q}) = \chi^{(0),DMFT}_{\omega}(\vc{q}) + g_{\omega}^2$, so that 
$\eta_{\omega}(\vc{q}) = -g_{\omega}^{-1} \left[\chi^{(0),DMFT}_{\omega}(\vc{q})\right]^{-1} \tilde \chi^{(0)}_{\omega}(\vc{q})$. In addition, the following relation holds: 
\begin{equation}
(1+g_{\omega}\Sigma_{\omega})(1+g_{\omega}(\Sigma_{\omega} - U)) = g_{\omega}^2  (K^0_{\omega}K^1_{\omega})^{-1} = \Lambda_{\omega}^{-1}.
\end{equation}
Eq. (\ref{S:B_DMFT}) can be rewritten as
\begin{equation}\label{S:B_DMFT_2}
B^{\mathrm{DMFT}}(\vc{q},T)=\sum_{\omega}\frac{-w_0 w_1 U^2 \Lambda_{\omega}^2 \tilde \chi^{(0)}_{\omega}(\vc{q}) g_{\omega}^2 }
{-(1+g_{\omega}(2\Sigma_{\omega}-U))\Lambda_{\omega}\tilde \chi^{(0)}_{\omega}(\vc{q})  + \underbrace{\tilde \chi^{(0)}_{\omega}(\vc{q}) - g_{\omega}^2}_{\chi^{(0),DMFT}_{\omega}(\vc{q})}
}.
\end{equation}
For the denominator we use
\begin{multline*}
(1+g_{\omega}(2\Sigma_{\omega}-U))\Lambda_{\omega} - 1 = g_{\omega}^{-2} ( w_0 (K^0_{\omega})^2 + w_1 (K^1_{\omega})^2) - 1 = g_{\omega}^{-2} (g_{\omega}^2 + w_0 w_1 (K^1_{\omega} - K^0_{\omega})^2) - 1 = w_0 w_1  U^2 \Lambda_{\omega}^2 g_{\omega}^{2}.
\end{multline*}
Thus
\begin{equation}
B^{\mathrm{DMFT}}(\vc{q},T)=\sum_{\omega}\frac{w_0 w_1 U^2 \Lambda_{\omega}^2 \tilde \chi^{(0)}_{\omega}(\vc{q})}
{1 + w_0 w_1  U^2 \Lambda_{\omega}^2 \tilde \chi^{(0)}_{\omega}(\vc{q})  },
\end{equation}
which equals $B(\vc{q},T)$ introduced in the previous section. 

The static charge susceptibility of conduction electrons within our approach is obtained as $\chi_{cc}^{\mathrm{DMFT}}(q,T) = T\sum_\omega \chi^{(0),\mathrm{DMFT}}_{\omega}(\vc{q}) + T \sum_{\omega\omega'} \chi^{(0),\mathrm{DMFT}}_{\omega}(\vc{q}) \tilde \Gamma_{\omega\omega'}(\vc{q})\chi^{(0),\mathrm{DMFT}}_{\omega}(\vc{q}) $. The equivalence of this formula and the expression for $\chi^{cc}(\vc{q})$ in Table I of Ref.~\citesup{supFreericks2003} has also been checked numerically.
\section{S4. Comparison between ladder and second-order dual fermion approximations}
\begin{figure}[h]
\includegraphics[width=0.8\columnwidth]{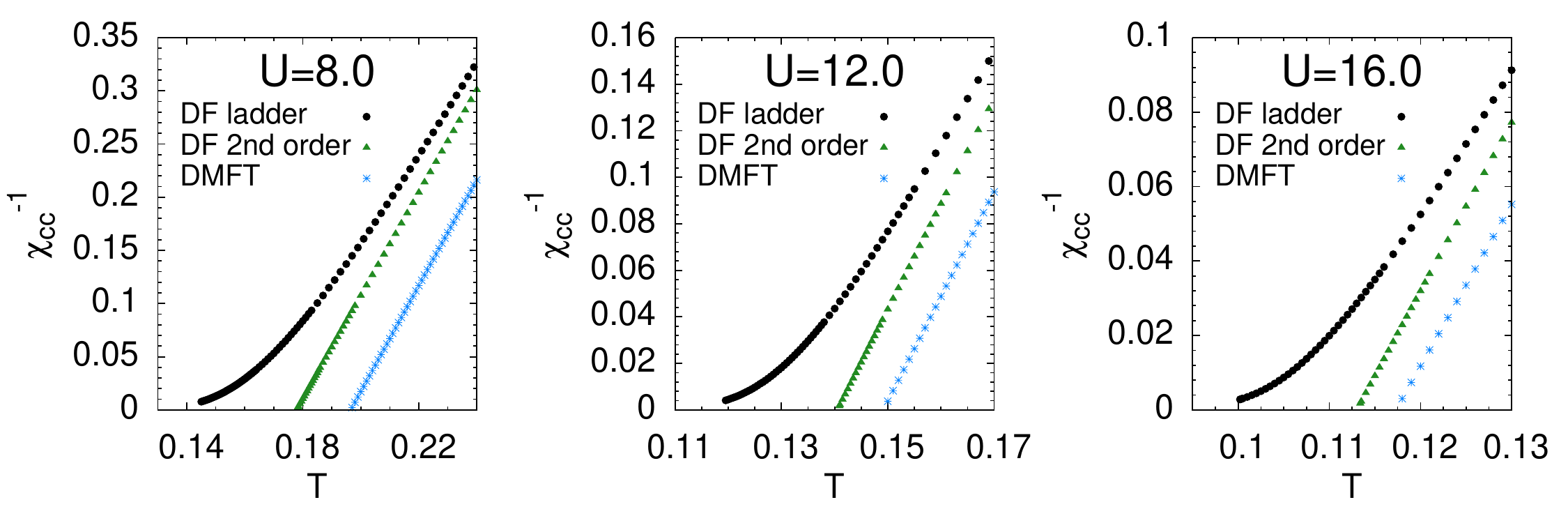}
\caption{Inverse of static c-electron charge susceptibility in 2$d$ as a function of $T$ at $U=8,12,16$ obtained within DMFT, ladder dual fermion approximation (\ref{S:bs_static}) and the second order dual fermion approximation (Eq. (\ref{S:bs_static}) truncated at the first iteration). The self-consistent evaluation of the second order dual fermion correction changes the amplitude and, consequently, $T_c$, but not the critical exponent as compared to the DMFT. The self-consistent summation of the dual fermion ladder results in the non mean-field critical exponent.}
\end{figure}


%

\newpage\end{widetext}
\end{document}